\documentclass[onecolumn]{aastex63}
\usepackage{amsmath}

\def\apj{{ApJ,~}}
\def\apjl{{ApJL,~}}
\def\mnras{{ MNRAS,~}}

\def\prd{{ Phys. Rev. D.,~}}
\def\prl{{ Phys. Rev. Lett.,~}}
\def\etal{{\it et al.~}}

\begin{document}
\title{Strong post-merger gravitational radiation of GW170817-like events}
\author{Yi-Zhong Fan}
\author{Jin-Liang Jiang}
\author{Shao-Peng Tang}
\author{Zhi-Ping Jin}
\author{Da-Ming Wei}
\affil{Key Laboratory of Dark Matter and Space Astronomy, Purple Mountain Observatory, Chinese Academy of Sciences, Nanjing 210023, China.}
\affil{School of Astronomy and Space Science, University of Science and Technology of China, Hefei, Anhui 230026, China.}
\email{yzfan@pmo.ac.cn (YZF) and dmwei@pmo.ac.cn (DMW)}

\begin{abstract}
The post-merger gravitational wave (GW) radiation of the remnant formed in the binary neutron star (BNS) coalescence has not been directly measured, yet. We show in this work that the properties of the BNS involved in GW170817, additionally constrained by PSR J0030+0451, the lower limit on the maximum gravitational mass of non-rotating neutron star (NS) and some nuclear data, are in favor of strong post-merger GW radiation. This conclusion applies to the mergers of Galactic BNS systems as well. Significant post-merger GW radiation is also preferred to improve the consistency between the maximum gravitational mass of the non-rotating NS inferred from GW170817/GRB170817A/AT2017gfo and the latest mass measurements of pulsars. The prominent post-merger gravitational radiation of GW170817-like events are expected to be detectable by advanced LIGO/Virgo detectors in the next decade and then shed valuable lights on the properties of the matter in the extremely high density.
\end{abstract}

\section{Introduction}
The mergers of binary neutron stars (BNSs) are one of the prime targets for the second generation gravitational wave (GW) detectors such as the advanced LIGO/Virgo and KAGRA \citep{Abbott2018}. During the inspiral the dynamics of neutron stars (NSs) is well described and the gravitational waveforms increase continually in both amplitude and frequency. After the merger the waveforms reflect the oscillations of the formed remnants (either black holes or supramassive NSs) and are much more complicated (in the case of the black hole formation, the waveforms terminate with the ringdown signal). The inspiral signal has a duration of tens of seconds (or even longer) and a low frequency (up to $\sim 1$ kHz), which is within the sensitive region of the advanced LIGO/Virgo and KAGRA detectors. The ringdown signal for black holes formed in BNS mergers is instead at frequencies of quite a few kHz, which is usually unmeasurable by the second generation detectors unless the sources are extremely close. The signals from the pre-collapse remnants are at frequencies lower than the ringdown, but still so high that a detection is challenging \citep[see][for a recent review]{Baiotti2019}. Such post-merger gravitational waves, anyhow, carry fundamental information on the equation of state (EoS) of the ultra-dense matter as well as the fate of the remnant formed in the BNS merger, and the interest in catching such a signal with the upgrading second generation gravitational detectors is growing. Important progresses have been achieved in the numerical simulations of the post-merger gravitational radiation \citep[e.g.,][]{Xing1994,Ruffert1996,Shibata2000,Damour2010,Hotokezaka2013,Bernuzzi2014,Bauswein2015,Bernuzzi2015,Zappa2018,Most2019,Bauswein2019,Baiotti2019} and dedicated efforts have been made to develop new data analysis methods \citep[e.g.][]{Clark2016,Yang2018}. An interesting finding of the numerical simulations is that the post-merger gravitational waves carry away in total about $0.8-2.5\%$ of mass-energy of BNS system, depending on the properties of the NSs as well as the EoS of the dense matter \citep{Bernuzzi2016,Zappa2018}.

On 2017 August 17, the advanced LIGO/Virgo discovered the gravitational-wave signal (i.e., GW170817) from the coalescence of a pair of NSs \citep{Abbott2017}. Very recently, \citet{Abbott2020} reported the detection of a new GW event GW190425 that involves at least one NS \citep{Han2020}. In comparison to GW170817, GW190425 was just detected by LIGO-Livingston and has a much lower signal-to-noise ratio (SNR). The dedicated search in the data of GW170817 found no signal from the post-merger remnant \citep{Abbott2017b}. In this work we evaluate its amount of post-merger GW radiation in two other ways. We find that the properties of the BNSs involved in GW170817, additionally constrained by PSR J0030+0451, the maximum gravitational mass of the non-rotating NS ($M_{\rm TOV}$) and some nuclear data, are in favor of efficient post-merger GW radiation. We further show that the strong post-merger GW radiation is preferred in improving the consistency among the $M_{\rm TOV}$ inferred from GW170817/GRB170817A/AT2017gfo and the latest mass measurements of pulsars. These two independent pieces of evidence are encouraging and the detection prospect of the post-merger GW signals from GW170817-like events in the full sensitivity run of LIGO/Virgo/KAGRA is found to be promising.

\begin{figure}[!ht]
    \begin{center}
    \includegraphics[width=0.95\linewidth, height=0.7\linewidth]{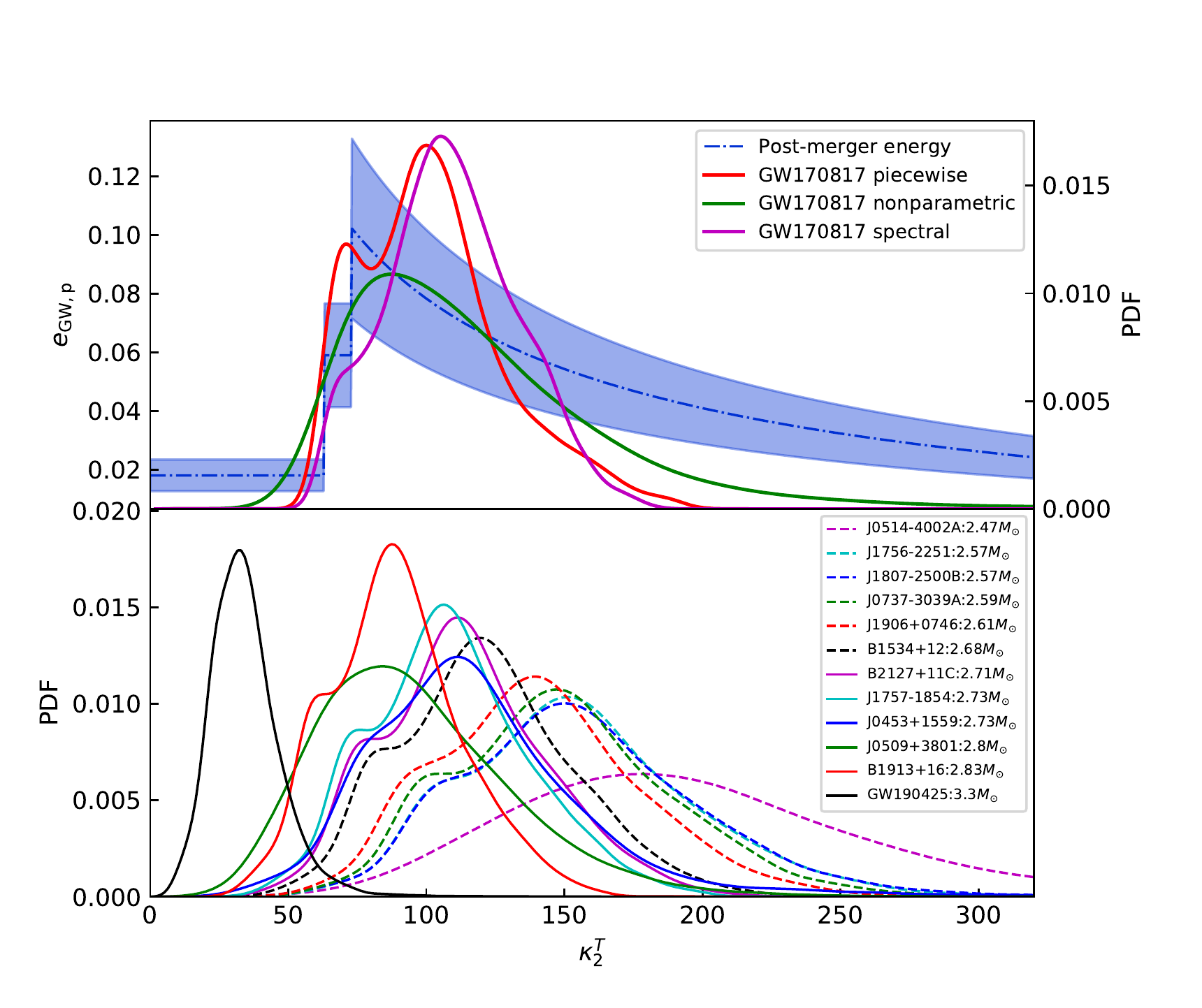}
    \end{center}
    \caption{Upper panel: the $\kappa_{2}^{T}$ distribution of GW170817. { The solid red (magenta) line represents the probability distribution function (PDF) of $\kappa_{2}^{T}$ of GW170817 directly evaluated from the posterior samples of the piecewise (spectral) method of \citet{Jiang2020}, while the solid green line is for the $\kappa_{2}^{T}$ evaluated from nonparametric result of PSRs+GWs+x-ray/Riley case in \citet{Landry2020}}. The dark blue dash-dotted line shows the best fit relation between $\kappa_{2}^{T}$ and the reduced GW energy emitted in the post-merger phase, { and the shaded blue region shows the corresponding $1\sigma$ fitting error}. Lower panel: the $\kappa_{2}^{T}$ distribution of GW190425 and some Galactic BNS systems that will merge in the Hubble time scale and have accurate mass measurements of individual components. The $M_{\rm tot}$ of each system \citep{Alsing2018,Lynch2018,Ridolfi2019,Abbott2020,Shao2020b} is marked. { Note that the PDF value of GW190425 has been reduced to half for clarity.}}
    \label{fig:k2t}
\end{figure}

\begin{figure}[!ht]
    \begin{center}
    \includegraphics[width=0.9\linewidth, height=0.65\linewidth]{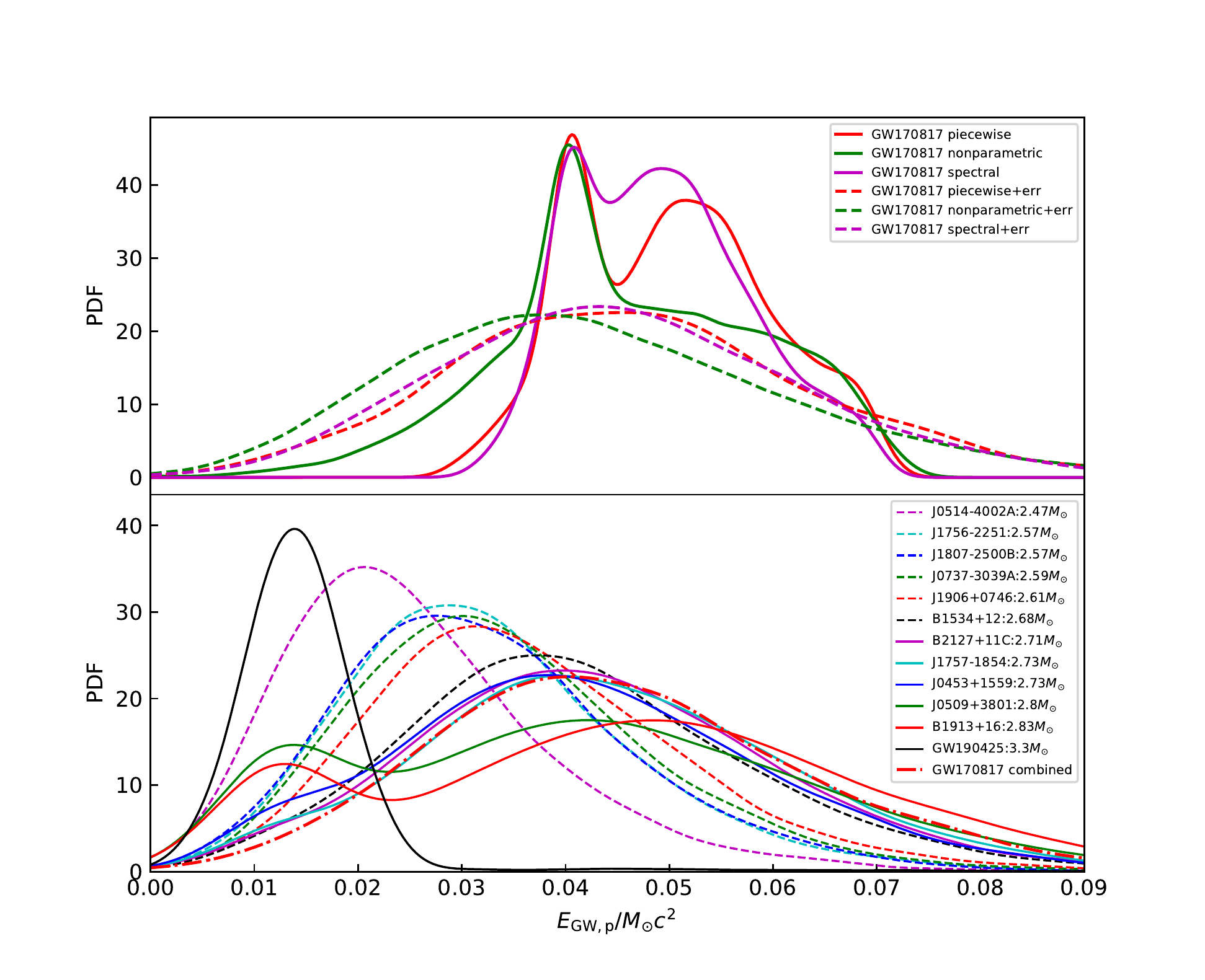}
    \end{center}
    \caption{ Estimated GW energy emitted during the whole post-merger phase. In the upper panel, the solid lines represent the estimated post-merger energy of GW170817 if the fitting error of eq.(\ref{eq:fit_func}) is not considered, while the dashed lines show the ones including the fitting error. In the lower panel, the $E_{\rm GW,p}$ of all the BNS sources considered in this work are shown, which have taken into account the joint posteriors of three different kinds of EoS parametrization method and the fitting error of eq.(\ref{eq:fit_func}). For the  two specific GW events, the error considered joint result of GW170817 is shown with the thick red dash-doted line, while the GW190425 is shown in the solid black line and its PDF is halved for clarify.}
    \label{fig:egw}
\end{figure}

\section{Strong post-merger gravitational radiation of GW170817}
\subsection{Prominent post-merger gravitational radiation anticipated for the properties of the BNSs of GW170817}
So far, the only way to theoretically quantify the radiated GW energy is to perform numerical relativity simulations. It turns out that $\kappa_{\rm 2}^{T}$, which parameterizes at leading-order the tidal interactions in the general-relativistic 2-body Hamiltonian, waveform's phase and amplitude \citep{Damour2012}, plays a very important role in estimating the GW radiation \citep{Bernuzzi2015,Bernuzzi2016}. The parameter $\kappa_{\rm 2}^{T}$ for a BNS system takes the form of
$\kappa_{2}^{T} = 3{(M_{\rm A}^{4}M_{\rm B}\Lambda_{\rm A}+M_{\rm B}^{4}M_{\rm A}\Lambda_{\rm B})}/{(M_{\rm A}+M_{\rm B})^5}$,
where $M$ is the gravitational mass, and $\Lambda$ is the dimensionless tidal deformability of a NS, which is related to the quadrupole Love number $k_{2}$ and the radius $R$ of the NS by $\Lambda=2/3(Rc^2/GM)^5k_{2}$ ($c$ is the speed of light and $G$ is the gravitational constant) \citep{Damour2010}. In general, the larger energy emissions correspond to smaller values of $\kappa_{2}^{T}$, which, in turn, gets smaller values for larger masses, more compact NSs and softer EoS. Using a large set of numerical relativity simulations with different binary parameters and input physics, \citet{Zappa2018} has found an empirical relation between $\kappa_{2}^{T}$ of the BNS system and the reduced gravitational-wave energy $e_{\rm gw,p}=E_{\rm GW,p}/(M_{\rm tot}\nu c^2)$ emitted in the post-merger phase, { which reads
\begin{equation}
e_{\rm GW,p}( \kappa^T_2 ) =
        \begin{cases}
            0.02  & \kappa^T_2 \lesssim 63 \\
                - & 63 \lesssim  \kappa_2^T \lesssim 73 \\
            a(\kappa^T_2)^{-\frac{7}{10}} + b & 73\lesssim \kappa^T_2 \lesssim 458 \\
            c \kappa^T_2 +d  & \kappa^T_2 \gtrsim 458,
        \end{cases}
\label{eq:fit_func}
\end{equation}
where $M_{\rm tot}$ is the total gravitational mass of the system, and $\nu=M_{\rm A}M_{\rm B}/M_{\rm tot}^{2}$ is the symmetric mass ratio. And the best fit values are $a = 2.44$, $b = -0.019$, $c =-5.1 \times 10^{-5}$, and $d = 0.038$ (please see the technical note at https://dcc.ligo.org/T1800417/public/ for the details)}. Their result is shown by the dash-dotted line in Fig.\ref{fig:k2t} and the left low $E_{\rm GW,p}$ region represents the prompt black hole formation scenario. { We have incorporated the non-negligible fitting error of this formula (see the upper panel of Fig.\ref{fig:k2t}) in a Monte Carlo way by using the distribution of the residual as done in the \citet{2019PhRvD..99l3026K}, and the broadened $E_{\rm GW,p}$ results are shown in the upper panel of Fig.\ref{fig:egw}.}

In this work, the parameter ranges are for 68\% credibility interval unless specifically mentioned. With eq.(\ref{eq:fit_func}), the amount of the post-merger gravitational radiation can be reasonably/qualitatively evaluated as long as $\kappa_{2}^{T}$ (i.e., the EoS) is known. However, various EoS models have been proposed in the literature and it is not possible to be uniquely determined even in the foreseeable future. Fortunately, under the reasonable assumption that all NSs follow the same EoS, their properties can be jointly/reliably constrained with the nuclear data, the GW data, the measured masses and the estimated radii of some NSs \citep[e.g.,][]{Lattimer2016,Tews2017,Abbott2017,Most2018,Landry2019}. The masses of NSs in some binary systems have been accurately measured and there is a robust lower limit on $M_{\rm TOV}\geq 2M_\odot$ \citep{Cromartie2020,Kandel2020}. The radii of NSs, however, usually are just evaluated indirectly and suffer from large systematical uncertainties. Thanks to the successful performance of the Neutron Star Interior Composition Explorer ({\it NICER}), the situation has changed and very recently the first-ever accurate measurement of mass and radius together for PSR J0030+0451, a nearby isolated quickly rotating NS, has been achieved \citep{Riley2019,Miller2019}, which favor a stiffer EoS than the data of GW170817. Hence, GW170817, PSR J0030+0451, some nuclear data as well as the lower limit on $M_{\rm TOV}$ can be combined to reliably constrain the EoS as well as the bulk properties of NSs. This can be done either in the EoS parameterizing methods \citep{Raaijmakers2019,Jiang2020} or the non-parametric approach \citep{Landry2020,Essick2020}, and the results are well consistent with each other. Here we directly adopt posterior samples of $\{M_{\rm A}, M_{\rm B}, \Lambda_{\rm A}, \Lambda_{\rm B}\}$ obtained in \citet{Jiang2020} to calculate the $\kappa_{2}^{T}$ for GW170817. Note that the region of $\kappa_{2}^{T}<63$ represents the prompt black hole formation, which is irrelevant to GW170817 because of the delayed collapse of the remnant \citep{Metzger2019}. { So we neglect the posterior samples that give $\kappa_{2}^{T}<63$ for GW170817. At $90\%$ credible level,} for the piecewise polytropic expansion method we have $\kappa_{2}^{T}=100^{+54}_{-34}$, while for the spectral decomposition method we have $\kappa_{2}^{T}=108^{+41}_{-38}$. { We also adopt the method described in \citet{2019PhRvD..99l3026K} to evaluate the nonparametric posterior of PSRs+GWs+x-ray/Riley case in \citet{Landry2020} and get $\kappa_{2}^{T}=107^{+103}_{-40}$ for GW1708017. The incorporation of the strong phase transition possibility by \citet{Landry2020} favors lower $k_2^T$ than ours (note that the $\kappa_{2}^{T}<63$ region is excluded), but has an overall agreement with the piecewise result and the spectral result (as shown in upper panel of Figure \ref{fig:k2t}). For this reason we combine equal sample of $k_2^T$ calculated from these three different parametrization methods to perform all the calculations in this work unless specially specified}. Clearly, the inferred $\kappa_{2}^{T}$ is well within the region that predicts the very prominent post-merger GW radiation, which is rather encouraging. { The post-merger energy is then estimated to be $E_{\rm GW,p} = 0.041^{+0.036}_{-0.026} M_{\odot} c^2 $ ($90\%$ confidence level) when the fitting error is considered for the combined posterior (see upper panel of Fig.\ref{fig:egw}). For B1534+12, B2127+11C, J1757-1854, and J0453+1559, we expect that they would emit almost the same amount of energy as GW170817 in the post-merger phase. This is understandable considering the comparable total mass of these systems. For lighter BNS systems like J0514-4002A, the expected post-merger energy is relatively small because of the large $\kappa_{2}^{T}$. While for heavier BNS systems like GW190425, we predict prompt collapse scenario and thus emit a small amount of energy in the post-merger phase (see the lower panels of Fig.\ref{fig:k2t} and Fig.\ref{fig:egw}).} To our knowledge, this is the {\it first} time to combine the intriguing numerical finding of \citet{Zappa2018} with the EoS constrained with the multi-messenger information of NSs and then demonstrate that GW170817 is likely the most efficient post-merger GW emitter { among the observed BNS GW events}.

\begin{figure}[!ht]
    \begin{center}
    \includegraphics[width=0.8\linewidth]{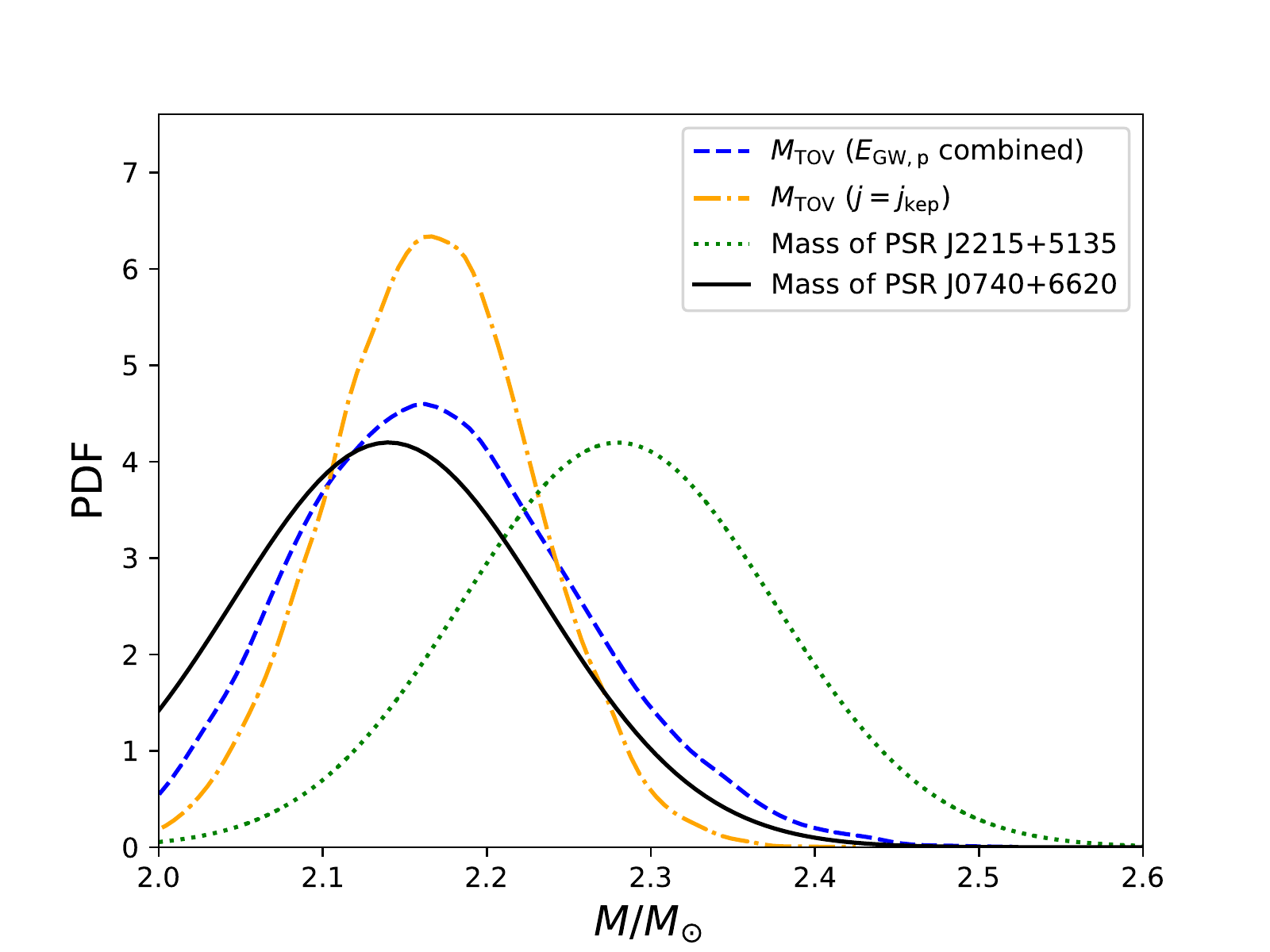}
    \end{center}
    \caption{The probability distribution functions of $M_{\rm TOV}$ { (in the cases of $j=j_{\rm kep}$ and  combined $E_{\rm GW,p}$)} inferred from GW170817/GRB 170817A/AT2017gfo and the mass distributions of PSR 0740+6620 \citep{Cromartie2020} and PSR J2215+5135 \citep{Kandel2020}.}
    \label{fig:mass-distribution}
\end{figure}

\subsection{Significant post-merger gravitational radiation: { indication} from the electromagnetic data}
The collapse time of the supramassive NS formed in GW170817 has been inferred to be $t_{\rm c}=0.98\pm 0.31$ s \citep{Gill2019},
when the uniform rapid rotation of the NS is insufficient to support against collapse. This imposes a constraint
on the $M_{\rm TOV}$ and the most-widely quoted limit is $\leq 2.17M_\odot$ \citep[e.g.,][]{Margalit2017,Shibata2017,Rezzolla2018}.

Very recently, \citet{Shao2020} derived an empirical relation among the critical total mass of BNSs ($M_{\rm tot,c}$), the mass and compactness (denoted by $\zeta_{\rm TOV}\equiv GM_{\rm TOV}/R_{\rm TOV}c^{2}$) of NS in the non-rotation maximum equilibrium configuration, the dimensionless angular momentum of remnant at the onset of collapse ($j_{\rm c}$) and the total mass lost apart from the remnant core ($m_{\rm loss}$), which reads
\begin{equation}
M_{\rm tot,c}\approx M_{\rm TOV}\left[1+0.122({j_{\rm c}}/{j_{\rm kep}})^2+0.040\left({j_{\rm c}}/{j_{\rm kep}}\right)^4\right](0.798+0.971\zeta_{\rm TOV})(1-0.091~M_\odot^{-1}~m_{\rm loss})+m_{\rm loss},
\label{eq:j-general}
\end{equation}
where $j_{\rm kep}\approx 1.24 \zeta_{\rm TOV}^{0.5}$ is the dimensionless angular momentum of NS spinning at Keplerian angular velocity. Supposing that the GW radiation just carried away a few percents solar mass energy in the post-merger phase, the pre-collapse remnant is expected to rotate at the mass-shedding limit \citep{Zappa2018}, as assumed in \citet{Margalit2017}. In such a case, we have $M_{\rm tot,c}\approx 1.162 M_{\rm TOV}(0.798+0.971\zeta_{\rm TOV})(1-0.091~M_\odot^{-1}~m_{\rm loss})+m_{\rm loss}$. At the collapse time ($t_{\rm c}\sim 1$s) of the central remnant of GW170817, $m_{\rm loss}$ consists of two parts, one is the kilonova/macronova outflow with a mass of $M_{\rm ej}\approx 0.05\pm0.01M_\odot$ \citep{Pian2017}, and the other is the accretion torus with a mass of $0.015-0.134M_\odot$ (90\% confidence level; and the most plausible value is $0.035M_\odot$, as found in the GRB 170817A/afterglow modeling by \citealt{Wang2019}). So far, $\zeta_{\rm TOV}$ is still not directly measurable and here we adopt the joint constraints set in \citet{Jiang2020}. Under the assumption of that the pre-collapse remnant was a supramassive NS supported by the rapid uniform rotation (i.e., $M_{\rm tot,c}=M_{\rm tot}$), we can evaluate $M_{\rm TOV}=2.16^{+0.06}_{-0.06}M_{\odot}$ (the 68\% credibility interval includes also the uncertainties of the EoS insensitive relationships adopted in \citet{Shao2020}) and the probability distribution is shown in Fig.\ref{fig:mass-distribution}. So far, the most massive NS is widely believed to be PSR J0740+6620, which has a mass of $2.14^{+0.10}_{-0.09}M_\odot$ \citep[][]{Cromartie2020}. PSR J2215+5135 may be more massive \citep[$M=2.28\pm 0.10M_\odot$;][]{Kandel2020} while the measurement method is not as direct/widely-accepted as that of PSR J0740+6620. $M_{\rm TOV}$ should be larger than the gravitational mass of any stable slowly-rotating cold NSs. For the inferred $M_{\rm TOV}(j=j_{\rm Kep})$, such a request is satisfied for PSR J0740+6620 but mildly violated for PSR J2215+5135.

The powerful GW radiation effectively carries away the angular momentum of the merger formed supramassive NS \citep{Shibata2019}. With a stronger GW radiation, the supramassive NS will rotate slower and a higher $M_{\rm TOV}$ is needed to support against the collapse \citep[e.g.][which is also evident in eq.(\ref{eq:j-general})]{Fan2013,Breu2016}. { Note that the inferred $E_{\rm GW,p}$ has a wide distribution (see Fig.\ref{fig:egw}) and the corresponding constraint on $M_{\rm TOV}$ will be modified in comparison to the case of $j=j_{\rm kep}$.} We therefore fully reproduce our calculation made in \citet{Shao2020}, adopting the $E_{\rm GW,p}$ found in Fig.\ref{fig:egw} (the combined case) and obtain { $M_{\rm TOV}=2.17^{+0.09}_{-0.09}M_{\odot}$ ($2.17^{+0.15}_{-0.12}M_\odot$ for the 90\% credibility; With the fixed $E_{\rm GW,p}=0.07M_\odot c^{2}$ we will yield $M_{\rm TOV}=2.23^{+0.12}_{-0.11}M_{\odot}$, which is similar to the mass of PSR J2215+5135)}. Intriguingly, \citet{Landry2020} found $M_{\rm TOV}=2.22^{+0.30}_{-0.20}M_\odot$ (90\% credibility) in the non-parametric constraints of NS matter with gravitational and pulsar observations \citep[very similar values have also been reported in][]{Essick2020}. The consistency between our results and those independently found in the non-parametric constraints of NS matter is encouraging and can be taken as an additional support of our current approach. {  As shown in Fig.\ref{fig:mass-distribution}, the combined $E_{\rm GW,p}$ case has a higher possibility to have $M_{\rm TOV}\geq 2.3 M_\odot$ than the case of $j=j_{\rm kep}$ because of the range extending to $E_{\rm GW,p}\geq 0.07M_\odot c^{2}$. If NSs as massive as $\approx 2.3M_\odot$ have been accurately measured (say, the mass of PSR J2215+5135 has been firmly confirmed) in the future, $E_{\rm GW,p}\approx 0.07M_\odot c^2$ (i.e., the post-merger GW radiation of GW170817 reaches the most promising range) will be needed if our current understanding of GW170817/GRB 170817A/AT2017gfo is robust, unless the high temperature effect has played a key role in softening the EoS and then triggering the collapse \citep{Shao2020b}.}

\begin{figure}[!ht]
    \begin{center}
    \includegraphics[width=0.8\linewidth]{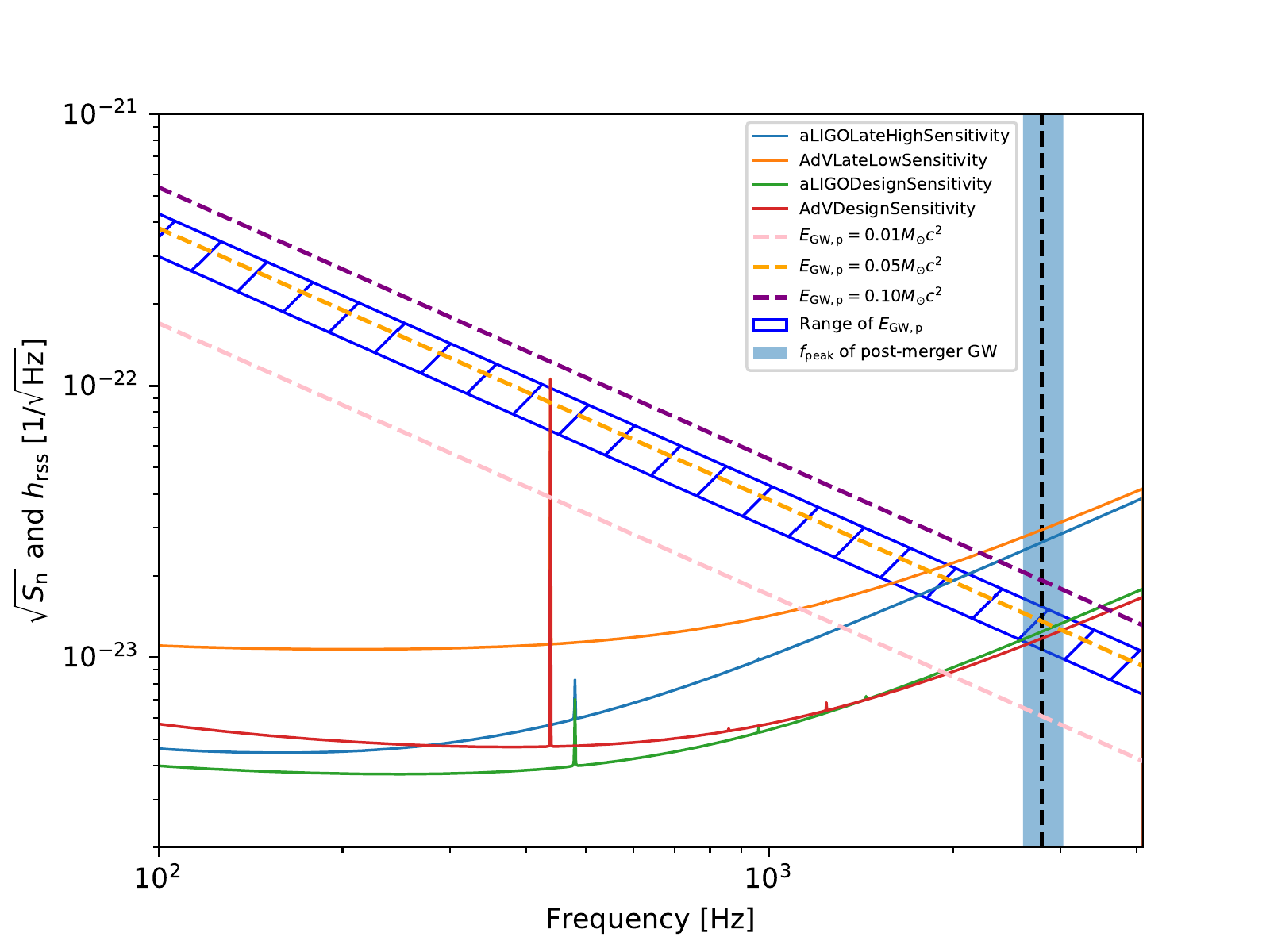}
    \end{center}
    \caption{The detection prospect of post-merger GW radiation of GW170817-like events in the O4 and O5 (design sensitivity) runs of the advanced detectors of LIGO/Virgo (the sensitivity curves are adopted from \citep{Abbott2018} and https://dcc.ligo.org/public/0161/P1900218/002/SummaryForObservers.pdf). Peak frequency of the post-merger GW is estimated using the eqs.(7-8) of \citet{2019PhRvD.100d4047T} with the posterior data taken from \citet{Jiang2020}.}
    \label{fig:detection-expect}
\end{figure}

\section{Detection prospect of the post-merger gravitational radiation of BNS mergers}
GW170817 was detected by advanced LIGO/Virgo in their second observing run. Though the SNR of the event is high, the sensitivity at $2-4$ kHz is still insufficient to catch the post-merger GW radiation \citep{Abbott2017b}. The improvements of the sensitivities are underway\footnote{\url{https://dcc.ligo.org/public/0161/P1900218/002/SummaryForObservers.pdf}}. In comparison to the O2 run, the sensitivity of the O5 run is expected to increase by a factor of $\sim 3.3$. This enhancement is about 1.5 times better than the initial design sensitivity. Therefore the detection prospect is more promising than that previously estimated with the initial design sensitivity. In Fig.\ref{fig:detection-expect}, the dashed lines indicate the maximum $h_{\rm rss}$ possible for a narrow band GW signal with some fixed energy contents $E_{\rm GW,p}$, under the most optimistic assumption that the whole energy available after merger is radiated in GWs at a certain frequency \citep[see also][]{Abbott2017b}, where the distance of the source is assumed to be the same as GW170817. Even so, the post-merger GW radiation of GW170817-like events is unlikely to be measurable in the upcoming O4 run for the strongest post-merger GW radiation with $E_{\rm GW,p}\sim 0.1M_\odot c^{2}$ (see the purple dashed line in Fig.\ref{fig:detection-expect}). A positive detection, anyhow, is plausible in the O5 observing run when the detectors reach their current design sensitivities for two good reasons. First, with a BNS merger rate of $\sim 10^{3}~{\rm Gpc^{-3}~yr^{-1}}$ (note that LIGO-India is expect to join in 2025, which will increase not only the total SNR of the event but also the factor of the duty cycle), we anticipate the detection of an event as close as $\sim 40$ Mpc in the next decade. Second, as shown in the lower panel of Fig.\ref{fig:egw}, except for the lightest and the heaviest BNS systems known so far, the $e_{\rm GW,p}$ values are in favor of efficient post-merger radiation of $E_{\rm GW,p}\sim {\rm quite~a~few}\times 10^{-2}~M_\odot c^2$. { Thus the hatched area in Fig.\ref{fig:detection-expect} should be taken as the fiducial case.} Different from the case of GW170817, now we calculate $\kappa_2^{T}$ in a simplified way. \citet{Jiang2020} have obtained $\Lambda$ as a function of $M$ (see Fig.3a therein; the piecewise method and the spectral method). As long as the gravitational masses of the BNSs are known, it is straightforward to estimate $\kappa_2^{T}$. { While for the non-parametric result, the universal relations in \citet{2019PhRvD..99l3026K} and the posteriors in \citet{Landry2020} are used to calculated the $\kappa_2^{T}$}. In the lower panel of Fig.\ref{fig:k2t}, besides the Galactic BNS systems with accurately measured individual masses, we include GW190425, for which the mass information is taken from the website\footnote{\url{https://dcc.ligo.org/LIGO-P2000026/public}; the IMRPhenomDNRT low-spin case.}. \citet{Abbott2020} speculated the prompt formation of black hole for GW190425 and here we do find weak post-merger GW radiation of $E_{\rm GW,p}\sim 0.014^{+0.009}_{-0.008}M_\odot c^{2}$. In the literature, with the dedicated numerical simulations, people have found that for some EoS models the post-merger GW radiation could be detectable for the sources as close as GW170817 supposing the sensitivity can reach a factor of a few times the initial advanced LIGO/Virgo sensitivity \citep[e.g.][]{Clark2016,Torres-Rivas2019}. { Our estimates of $E_{\rm GW,p}$, however, are EoS-insensitive} and the detection prospect are consistent with these EoS-dependent numerical evaluation.

\section{Summary}
So far, two BNS merger events (GW170817 and GW190425) have been reported by the LIGO scientific collaboration and Virgo collaboration and the measurements are solely for the inspiral signals. The post-merger gravitational waves carry fundamental information on the EoS of the ultra-dense NS matter as well as the fate of the remnant formed in the BNS merger, and have attracted wide attention. The typical frequencies $\sim 2-4$ kHz of the post-merger signals however are out of the most sensitive region of the second generation GW detectors, which well explains the absence of such signals in current BNS merger events. Thanks to the rapid progresses made in the numerical simulations, an intriguing relation between $\kappa_2^{T}$ and the strength of the post-merger GW radiation has been suggested in the literature \citep[e.g.,][]{Bernuzzi2016,Zappa2018}. We have calculated $\kappa_2^{T}$ of the BNSs involved in GW170817, benefited with the further constraints from PSR J0030+0451, some nuclear data as well as the robust lower limit on the maximum gravitational mass of the non-rotating NSs. { With the distributions of $\kappa_2^{T}$ for three parameterization methods considered in this work and the numerical fitting relation, we transform the $\kappa_2^{T}$ to the $E_{\rm GW,p}$ which in favor of efficient post-merger GW radiation (see the upper panel of Fig.\ref{fig:egw}). Moreover, we also show that if in the future neutron stars as massive as $\approx 2.3M_\odot$ have been accurately measured, the post-merger GW radiation of GW170817 should be within the high end part of our inferred distribution (i.e., it is a very efficient post-merger GW radiator) unless new effects/assumptions have been introduced.} We finally show that for typical BNS systems, the post-merger GW radiation are expected to be strong (see the lower panel of Fig.\ref{fig:egw}). Together with a BNS merger rate of $\sim 10^{3}~{\rm Gpc^{-3}~yr^{-1}}$, in the O5 run of advanced LIGO/Virgo/KAGRA/LIGO-India, a few events as close as GW170817 are expected and their post-merger GW radiation are detectable. The successful detection would shed valuable lights on the physical properties of the ultra-dense matter.

\section*{Acknowledgments}
We thank the anonymous referee for the constructive suggestions. We are also grateful for the kind help from P. Landry and F. Zappa. This work was supported in part by NSFC under grants of No. 11525313 (i.e., Funds for Distinguished Young Scholars), No. 11921003 and No. 11773078, the Funds for Distinguished Young Scholars of Jiangsu Province (No. BK20180050), the Chinese Academy of Sciences via the Strategic Priority Research Program (Grant No. XDB23040000), Key Research Program of Frontier Sciences (No. QYZDJ-SSW-SYS024). This research has made use of data and software obtained from the Gravitational Wave Open Science Center \url{https://www.gw-openscience.org}, a service of LIGO Laboratory, the LIGO Scientific Collaboration and the Virgo Collaboration. LIGO is funded by the U.S. National Science Foundation. Virgo is funded by the French Centre National de Recherche Scientifique (CNRS), the Italian Istituto Nazionale della Fisica Nucleare (INFN) and the Dutch Nikhef, with contributions by Polish and Hungarian institutes.\\

\end{document}